# Revolutionizing Pharma: Unveiling the AI and LLM Trends in the Pharmaceutical Industry

Yu Han[1*] and Jingwen Tao[2]

[1]Department of Engineering Science, University of Oxford, Old Road Campus, Oxford, OX3 7QD, United Kingdom.
[2]Shanghai Mingdu Intelligent Cloud Computing Co., Ltd., No. 29, Lane 388, Shengrong Road, Shanghai, China.

*Corresponding author(s). E-mail(s): yu.han@eng.ox.ac.uk;

# 1 Abstract

This document offers a critical overview of the emerging trends and significant advancements in artificial intelligence (AI) within the pharmaceutical industry. Detailing its application across key operational areas, including research and development, animal testing, clinical trials, hospital clinical stages, production, regulatory affairs, quality control and other supporting areas, the paper categorically examines AI's role in each sector. Special emphasis is placed on cutting-edge AI technologies like machine learning algorithms and their contributions to various aspects of pharmaceutical operations. Through this comprehensive analysis, the paper highlights the transformative potential of AI in reshaping the pharmaceutical industry's future.

# 2 Introduction

The pharmaceutical industry plays a crucial role in global healthcare, driven by continuous innovation and a relentless pursuit of therapeutic advancement. It is a dynamic and multifaceted field that encompasses various critical aspects contributing to final product. At its core, "Research and Development" (R&D) serves as the driving force behind innovation and therapeutic advancement, which is an intricate process that involves the discovery of novel drug candidates, their meticulous evaluation, and subsequent clinical trials to ensure safety and efficacy. This stage represents the inception of groundbreaking treatments that have the potential to transform patient outcomes.[1] Complementing the R&D endeavor is the domain of "Manufacturing," where precise and standardized processes are meticulously executed. It is the critical role of manufacturing that ensures the reproducibility and quality of pharmaceutical products.[2] Furthermore, "Quality Control" procedures play a pivotal role in maintaining the highest standards throughout the manufacturing process.[3] Meanwhile, the pharmaceutical industry operates within a stringent regulatory framework to uphold safety and efficacy standards. Regulatory compliance is a cornerstone of the industry, with authorities such as the Food and Drug Administration (FDA) and the European Medicines Agency (EMA) overseeing product approvals. In parallel, various supporting functions, for example,"Supply Chain Optimization" ensure the timely delivery of pharmaceutical products to patients. The pharmaceutical industry, with its multifaceted dimensions and unwavering commitment to research, quality, regulation, and supply chain efficiency, stands as a beacon of innovation and a driving force in advancing healthcare on a global scale.

In this dynamic landscape, the emergence of Artificial Intelligence (AI) and Large Language Models (LLMs) marks a transformative era, offering unprecedented capabilities to reshape every aspect of the pharmaceutical industry.[4] In light of this, our research focuses on the application of artificial intelligence within the pharmaceutical industry. We aim to identify and analyze the various sectors of this industry where AI and LLMs have been implemented. Our study provides an abundance of information, offering a detailed examination of the specific areas where these technologies are being researched and utilized. Through this review, we seek to provide a comprehensive understanding of the current state and future prospects of AI and LLM applications, offering insights into how they are revolutionizing the pharmaceutical industry.

## 3 Methodology

Our research methodology employed a comprehensive approach to data collection, utilizing targeted search queries across major academic databases, including IEEE Xplore and Pubmed. The time frame for our search spanned from 2019 to 2024, chosen to capture the most recent and relevant advancements in AI within the pharmaceutical industry. Additionally, commercial databases were searched to ensure a broad coverage of the industry's latest developments.

The structure of our study meticulously segmented the pharmaceutical industry into distinct operational domains: research and development, manufacturing, quality control, regulatory affairs, clinical applications, and other supportive areas. This segmentation, depicted in Figure 1, was instrumental in facilitating an organized and detailed analysis of AI's diverse and multifaceted influence throughout the industry.

Our search strategy was methodically formulated, with each query being meticulously tailored to address specific aspects of each industry segment. For example, to probe into the utilization of AI in data analysis within the pharmaceutical domain, the search string **(pharmaceutical) AND ((artificial intelligence) OR (large language model)) AND (data analysis)** was employed. This targeted approach was replicated across other sectors, including laboratory data management and image recognition, to ensure a focused and relevant collation of data.

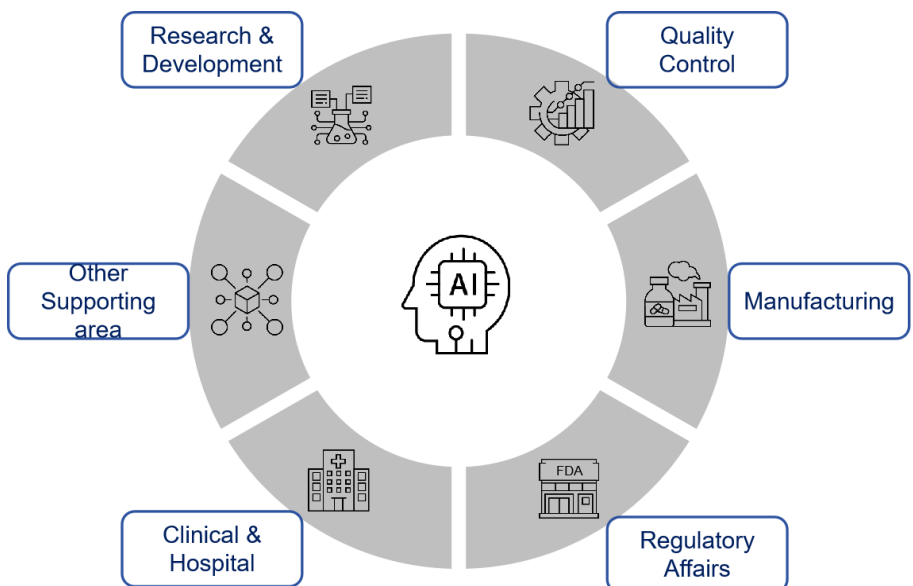

**Fig. 1**: Comprehensive Overview of Key Domains in the Pharmaceutical Industry

## 4 Results

The initial search yielded 7,402 papers, as shown in Table 1. These papers were first screened based on titles and abstracts, focusing on relevance and scope. This screening was conducted by multiple reviewers to ensure consistency. The initial distribution of research publications across the different segments of the pharmaceutical industry is illustrated in Figure 2. This visualization reveals a predominant focus on technological advancements in various areas. Notably, papers pertaining to Research and Development (R&D) outnumber those in other segments. Within this R&D category, studies concentrating on data analysis and drug discovery are particularly prominent, indicating a significant research emphasis in these areas. Additionally, the fields of regulatory affairs and document intelligence, as well as the integration of technology with AI-driven chatbots, emerge as significant areas of applied research.

Papers that passed this preliminary filter underwent a comprehensive full-text review. Here, we assessed each study's impact, methodological soundness, relevance to our topics, and innovation. This evaluation process ensured that our review encompassed the most significant and influential studies, providing a thorough overview of the current state and potential future of AI and LLMs in the pharmaceutical industry, as shown in Table 2. In the subsequent sections, we will present a detailed analysis of the selected papers.

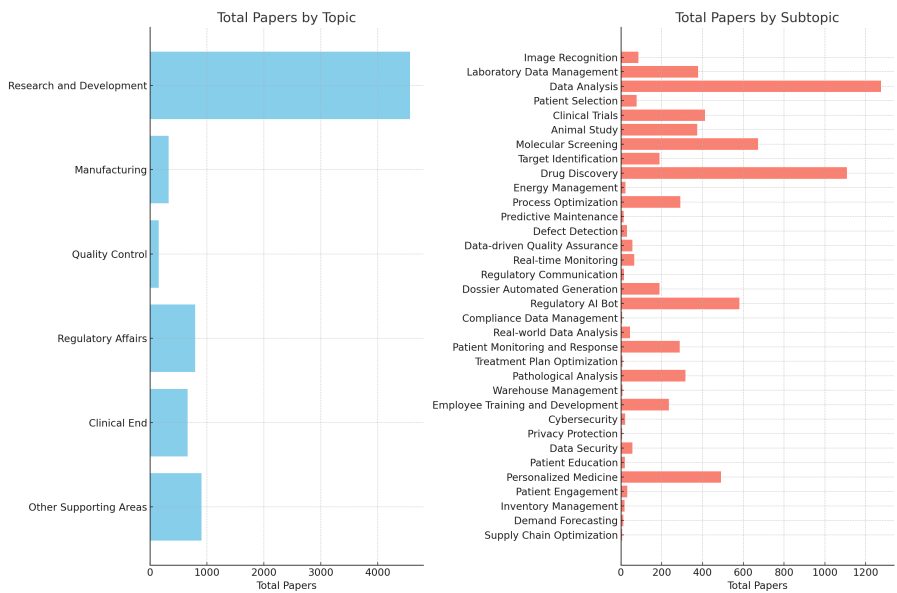

**Fig. 2**: Number of Publications Across Various Segments of the Pharmaceutical Industry

| Sub section | Application Area | Country | AI section | App. or Thry. | Citation |
| --- | --- | --- | --- | --- | --- |
| Drug Discovery | prediction of drug-target and drug–drug interactions | Australia | Graph-Based Approaches | Thry. | [5] |
| | Rare disease | USA | Graph-Based Approaches | Thry. | [6] |
| | Protein-ligand interaction | the Republic of Korea | Deep Learning and Neural Networks | App. | [7] |
| | Drug repurposing | Iran | Deep Learning and Neural Networks | App. | [8] |
| | drug formulation development | Canada | Machine learning | App. | [9] |
| Target Identification | Prediction protein targets of herbs treating Osteoporosis | China | Graph-Based Methods | App. | [10] |
| | Identification and Validation of LncRNA Signatures | China | Machine learning | App. | 11 |
| Animal Study | Network data of mouse gene | China | Machine learning | App. | [12] |
| | Gene of pig, monkey, rabbit and horse | China | disease modeling | App. | [13] |
| | Revealed high ADE associations for 2 drugs used in dogs and cats. | China | Machine learning | App. | [14] |
| Clinical Trials | Adopting AI in clinical trial Patient recruitment, endoscopic result | USA | Machine learning | App. | [15] |
| | Prediction of Clinical Trials Outcomes | United Arab Emirates | Natural language processing (NLP) | Thry. | [16] |
| | Optimizing Eligibility Screening for Clinical Trials | USA | Natural language processing (NLP) | Thry. | [17] |
| | ML application in phase I clinical trial | Italy | Machine learning | App. | [18] |
| | Generative artificial intelligence empowers digital twins | Germany | Deep Learning and Neural Networks | Thry. | [19] |
| | Simulates human organs on a chip | China | Machine learning | Thry. | [20] |
| Image recognition | Accurate diagnosis of colorectal cancer based on histopathology images | China | Deep Learning and Neural Networks | Thry. | [21] |
| | DeepCycle is a deep neural network able to reconstruct a cyclic cell cycle trajectory from unsegmented cell images | Germany | Deep Learning and Neural Networks | Thry. | [22] |
| Manufacturing | Process optimization in silico prediction | Australia | Machine learning | Thry. | [23] |
| | Predictive controller design for cell culture processes get maximized cell growth | USA | Machine learning | Thry. | [24] |
| Predictive Maintenance | Decrease the maintenance costs of the equipment. | Italy | Data Clustering and Frameworks: | App. | [25] |
| | Enforcing Traceability in Manufacturing Processes | Greece | Data Clustering and Frameworks | App. | [26] |
| | Drug Sales Prediction | India | Machine learning | App. | [27] |
| M-energy management | Intelligent control and management information system for optimizing drying process | Russia | Machine learning | Thry. | [28] |
| | Smart Healthcare and Safety Management | India | IoT and Miscellaneous Technologies | App. | [29] |
| Quality control | coating thickness measurement and defect recognition of film coated tablets | Hungary | Machine learning | App. | [30] |
| | Machine vision used in industrial product lines for quality control (QC) and quality assurance (QA). | Iran | Broader AI areas | App. | [31] |
| | video-assisted control system in a chemotherapy compounding unit | France | Machine learning | Thry | [32] |
| Defect detection | Product Defect Surveillance | Singapore | Machine learning | App. | [33] |
| | Internal Tablet Defect Detection | USA | Deep Learning and Neural Networks | App. | [34] |
| | Monitoring the powder bed fusion (PBF) processes | China | Machine learning | Thry | [35] |
| Regulatory Affairs | Integration of AI for digital transformation | India | Machine learning | App. | [36] |
| RA- compliance data management | Automated methods for calculating, monitoring and predicting compliance. | Italy | Machine learning | App. | [37] |
| | Automate routine work AI-Based Static and Dynamic Systems | Belgium | Machine learning | Thry. | [38] |
| RA - auto filing | LLM chatbots deployed in biomedical contexts | UK | Natural language processing (NLP) | Thry. | [39] |
| | Document classification and entity extraction of certain area documents | USA | Natural language processing (NLP) | Thry. | [40] |
| | Use NLP to harmonize Product information (PI) | Sweden | Natural language processing (NLP) | App. | [41] |
| Chatbot | Doctor Dialog agents (chatbots) with patients | India | Natural language processing (NLP) | App. | [42] |
| | Health Care Chatbots | Qatar | Natural language processing (NLP) | App. | [43] |
| | Surgery Chatbots | Turkey | Natural language processing (NLP) | App. | [44] |
| Clinical decision support | Analyze ECG data achieved comparable performance as doctors | USA | Deep Learning and Neural Networks | Thry. | [45] |

6| | | | | | |
|---|---|---|---|---|---|
| | APPRAISE-AI quantitative tool for evaluating the methodological rigor and clinical utility of AI models. | Canada | Broader AI areas | Thry. | 46 |
| | Cerebrospinal fluid (CSF) analysis based on a machine-learning-aided cross-reactive sensing array | China | Machine learning | Thry. | [47] |
| | Diabetes Management | Malaysia | Machine learning | App. | [48] |
| | Drug Combination Prediction | China | Machine learning | App. | [49] |
| Image recognition | Estimation of Visual Function | Japan | Deep Learning and Neural Networks | Thry. | [50] |
| | predicting longitudinal brain MRIs | USA | Broader AI areas | Thry. | [51] |
| Supply chain | Supply Chain Management | India | Deep Learning and Neural Networks | App. | [52] |
| | Block-chain in the medical supply chain optimization | India | Machine learning | Thry. | [53] |
| Data analysis | Extraction of medications and associated adverse drug events (ADEs) from clinical documents, | China | Natural language processing (NLP) | App. | [54] |
| Laboratory Management | AI analyze urine samples | Switzerland | Deep Learning and Neural Networks | App. | [55] |
| | Strategic choices faced by large pharmaceutical laboratories | Spain | Broader AI areas | Thry. | [56] |

**Table 2**: Curated Compilation of Research Publications: Domain Applications, Geographical Distribution, and Citation Metrics

An alluvium diagram, as shown in Fig 4 has been shown to visually represent the distribution of these papers, categorizing them based on the specific sections of the pharmaceutical industry, the country of origin, the domain of AI technology (explained as stated below), the application area, and the citation metrics of each paper. Papers were diligently sorted based on the technologies mentioned and organized into six primary Artificial Intelligence (AI) categories:

1. **Machine Learning Techniques:** This category includes a spectrum of algorithms such as Support Vector Machines (SVM), Reinforcement Learning, and other established machine learning methods.
2. **Deep Learning and Neural Networks:** This includes models, Convolutional Neural Networks (CNN), Transfer Learning, Digital Twins (DTs), and similar approaches that are based on neural network architectures.
3. **Natural Language Processing (NLP):** This segment encompasses all forms of NLP, including Large Language Models (LLMs).
4. **Graph-Based Approaches:** Entails methods that utilize network and knowledge graphs, along with other graph-related techniques.
5. **Data Clustering and Frameworks:** Covers specialized frameworks such as Density-Based Spatial Clustering of Applications with Noise (DBSCAN), Federated Learning Frameworks, and other related clustering or data framework technologies.
6. **IoT and Miscellaneous Technologies:** A general category for a range of technologies such as the Internet of Things (IoT) and any other technologies that do not distinctly fit into the other outlined categories.

Table 1: Total Records by Section and Subsection

| Main Section | Subsection | IEEE | Pubmed | Total Records |
|---|---|---|---|---|
| | Drug Discovery | 40 | 1,068 | 1,108 |
| | Target Identification | 3 | 186 | 189 |
| Research and Development | Molecular Screening | 9 | 663 | 672 |
| | Animal Study | 2 | 372 | 374 |
| | Clinical Trials | 9 | 403 | 412 |
| | Patient Selection | 3 | 74 | 77 |
| | Data Analysis | 86 | 1,189 | 1,275 |
| | Laboratory Data Management | 13 | 366 | 379 |
| | Image recognition | 20 | 66 | 86 |
| Manufacturing | Predictive Maintenance | 5 | 8 | 13 |
| | Process Optimization | 27 | 264 | 291 |
| | Energy Management | 8 | 15 | 23 |
| Quality Control | Real-time Monitoring | 7 | 58 | 65 |
| | Data-driven Quality Assurance | 1 | 55 | 56 |
| | Defect Detection | 3 | 27 | 30 |
| | Compliance Data Management | 1 | 5 | 6 |
| Regulatory Affairs | Regulatory AI Bot | 256 | 325 | 581 |
| | Dossier Automated Generation | 13 | 176 | 189 |
| | Regulatory Communication | 0 | 15 | 15 |
| | Pathological Analysis | 2 | 315 | 317 |
| Clinical end | Treatment Plan Optimization | 0 | 8 | 8 |
| | Patient Monitoring and Response | 32 | 257 | 289 |
| | Real-world Data Analysis | 3 | 42 | 45 |
| | Supply Chain Optimization | 4 | 2 | 6 |
| | Demand Forecasting | 2 | 10 | 12 |
| | Inventory Management | 3 | 15 | 18 |
| | Patient Engagement | 1 | 30 | 31 |
| Other Supporting Areas | Personalized Medicine | 25 | 465 | 490 |
| | Patient Education | 12 | 7 | 19 |
| | Data Security | 34 | 22 | 56 |
| | Privacy Protection | 2 | 4 | 6 |
| | Cybersecurity | 3 | 18 | 21 |
| | Employee Training and Development | 1 | 234 | 235 |
| | Warehouse Management | 3 | 5 | 8 |
| **Total** | | **633** | **6769** | **7402** |

The alluvial diagram provides a comprehensive visualization of the distribution and focus of the research papers we have analyzed. It elucidates the allocation of research efforts across various sectors within the pharmaceutical industry. Notably, the domain of Research and Development (R&D) emerges as the most extensively studied area, reflecting the industry's emphasis on innovation and advancement. Following that, there is a noticeable concentration of research in the Regulatory Affairs domain. This trend is indicative of the industry's growing focus on navigating complex regulatory landscapes, which is increasingly critical in today's globalized pharmaceutical market. Further analysis of the flow within the diagram towards practical applications. This underscores the application-oriented nature of this research, aimed at addressing real-world challenges in regulatory compliance and management. Then we could see substantial portion of this research in Regulatory Affairs is seen converging towards Natural Language Processing (NLP) technologies. The prominence of NLP in this context reflects the industry's adoption of advanced AI techniques to streamline regulatory processes, improve compliance efficiency, and facilitate effective communication within the regulatory framework. Progressing further into the analysis, we observe the predominance of Machine Learning as the most employed AI technology in the pharmaceutical industry, closely followed by Deep Learning applications. This trend highlights the significant role these technologies play in advancing pharmaceutical research and development. The final aspect of the diagram delineates the geographical distribution of the research. Predominantly, it is observed that China and the United States are at the forefront in terms of the volume of research output, following by other countries.

A world map visualization, as shown in Figure 4 is presented to illustrate the geographic distribution and thematic preferences of research conducted by scholars in different countries. This visual analysis aims to offer insights into the current global research trends in the field. We could see there is a pronounced concentration of Research and Development (R&D) efforts in China and the United States, highlighting these countries as key players in driving innovation and advancement within the pharmaceutical sector. Additionally, Russia and Australia emerge as notable hubs for research in manufacturing, indicating their significant contributions to this aspect of the industry. In Canada, there is a discernible emphasis on clinical phase research, reflecting the country's focused approach towards the development and testing of new therapeutic solutions. Meanwhile, India is identified as a prominent center for research in quality control. This trend underlines India's commitment to ensuring the safety, efficacy, and reliability of pharmaceutical products.

The comprehensive results of both research endeavors and published materials are meticulously delineated within the subsequent sections.

## 4.1 Research and Development

The burgeoning deployment of Artificial Intelligence (AI) and Language Learning Models (LLM) in the Research and Development (R&D) realm of the pharmaceutical industry is pivotal for both innovation and efficiency. This section delves into the array of AI and LLM methodologies applied in drug discovery, target identification, animal studies, clinical trials, and image recognition, as illuminated by contemporary research. This section comprehensively outlines the results of research and literature in the following subsections.

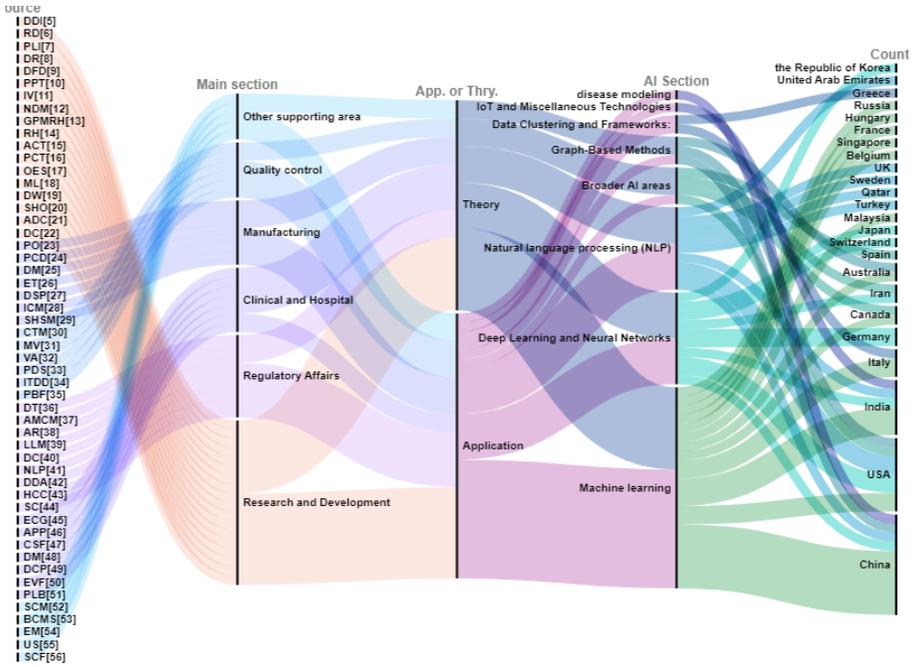

**Fig. 3**: Alluvial Analysis of Research Papers: Distribution by Country and Principal Domain of Study

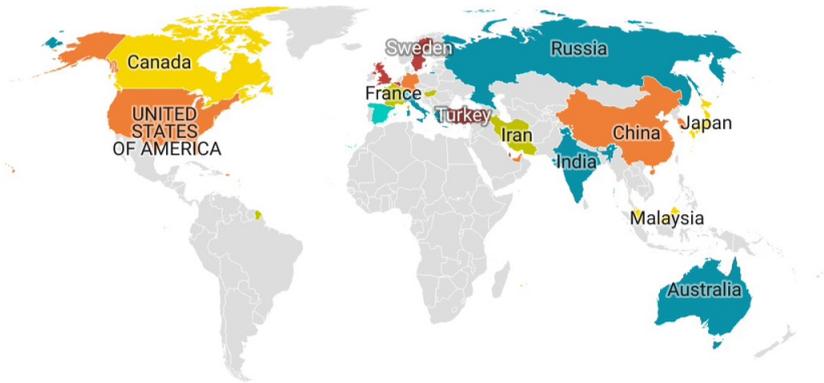

**Fig. 4**: World Map of Research Papers: Principal Domain of Study

### 4.1.1 Drug Discovery

AI's capability in predicting drug-target and drug-drug interactions is a cornerstone in the search for new drugs. Notably, network analysis leveraging graph theory has become a sophisticated AI approach for the prediction of molecular interactions. Models such as ResBiGAAT innovate further by integrating deep learning techniques with self-attention mechanisms to enhance protein-ligand binding predictions.[5] Additionally, convolutional neural networks are redefining drug re-purposing strategies by efficiently mining drug-disease association data. In activity of prediction of Drug-Target and Drug-Drug Interactions, graph theory has been instrumental in understanding the complex networks of biomedical data. By representing large datasets as networks, with nodes and edges symbolizing molecules and their interactions, AI models are able to predict potential drug-target and drug-drug interactions.[5] These link prediction models are particularly useful for identifying novel connections within the vast chemical space, expediting the discovery of potential therapeutic compounds. The IDDI-DNN, a deep neural network, integrates drug-disease association data to uncover new uses for established drugs.[8] This approach is particularly efficient in navigating the vast array of available pharmaceutical data, identifying potential drug repurposing opportunities with greater speed and accuracy. Besides, a novel deep learning framework, employs a deep Residual Bidirectional Gated Recurrent Unit coupled with a two-sided self-attention mechanism.[7]

Drug formulation is a critical process in the pharmaceutical industry, involving the combination of inert materials and excipients with active pharmaceutical ingredients (APIs) to create effective drug products. The goals of optimized drug formulation are multifaceted, aiming to enhance the drug's efficacy, prolong its therapeutic effects, minimize side effects, and extend the API's stability and shelf life, ultimately improving patient compliance. The choice of materials for formulation is diverse, including inert excipients like polymers, lipids, and surfactants, as well as other APIs. This allows for the creation of various delivery systems such as microparticles (MPs), nanoparticles (NPs), and multicomponent systems, which cater to different routes of administration and indication-specific requirements. These systems are then processed into final drug products, such as solid, liquid, or parenteral dosage forms. Although traditional formulation development methods have led to successful drug products, they are often time-consuming and may lack efficiency. Machine learning plays a crucial role in the formulation development phase, optimizing the composition and potential efficacy of pharmaceutical agents.

Typically, this process involves preparing and characterizing multiple candidate formulations through a comprehensive API-material matching process. These formulations are then evaluated and compared in terms of their performance against the API alone and competitor formulations, using a variety of in vitro and/or ex vivo assays.[9]

Knowledge graph is a sophisticated method for organizing and storing information that emphasizes the interconnectedness of various data elements. Essentially, it's a graphical depiction of knowledge, where entities, for example, objects, events, concepts, or individuals - are represented as nodes, and their interrelationships are illustrated as edges or links. Originally rooted in the field of computer science, knowledge graphs have increasingly found applications in the medical domain. It has been applied in constructing a dynamic map of disease-related knowledge, allows for the aggregation and analysis of fragmented data across various sources.[6] This method enhances the understanding of disease mechanisms and the identification of potential therapeutic targets, especially in area of rare disease with limited data availability.

In the crucial phase of target identification within pharmaceutical R&D, AI has emerged as a powerful tool, for example, AI applications in cheminformatics have revolutionized the way researchers approach traditional medicine. Tools such as Gephi, along with RDKit and SKlearn, are utilized to analyze the chemical composition of herbs used in treating osteoporosis.[10] Furthermore, the use of machine learning classifiers has been instrumental in the identification and validation of LncRNA signatures.[11] LncRNAs play a significant role in the regulation of various biological processes and are increasingly recognized for their importance in disease prognosis and therapy. The application of machine learning, especially Conditional Random Field (CRF)-like models, in analyzing LncRNA data sets provides a deeper understanding of their involvement in diseases such as clear cell renal cell carcinoma. This understanding is critical for the development of targeted therapies and for improving the outcomes of immunotherapy treatments.[11] The application of AI in target identification is proving to be invaluable in uncovering potential new therapies.

### 4.1.2 Animal Study

The application of AI in animal studies is revolutionizing the way pre-clinical research is conducted. This section examines the impact of AI algorithms on the study of animal genetics and the prediction of adverse drug events (ADE).

AI has been used in predictive modeling in genomics. Support Vector Machine (SVM) prediction models are being utilized to interpret complex genomic data. Scholars designed the MouseNet database incorporates these models to predict the mouse interactome and provide network-based interpretations of gene expressions.[12] This enables researchers to understand genetic interactions and their implications in disease states more accurately from animal study to all studies afterwards. Such predictive modeling is essential for identifying potential therapeutic targets and understanding disease pathology at the genetic level.

Gene editing techniques like ZFN, TALEN, and CRISPR/Cas9 are indispensable tools in modern biomedical research, and their application is enhanced by AI algorithms in disease modeling. By editing the genes of large animals such as pigs, monkeys, rabbits, and horses, researchers can model human diseases more effectively. AI assists in predicting the outcomes of these gene editing procedures, optimizing the models for more accurate representations of human diseases.[13] AI has also made significant advances in pharmacovigilance, particularly in analyzing and predicting ADEs. By employing various distance measures, such as Pearson, Spearman, cosine, Yule, and Euclidean distances, AI algorithms can detect associations between drugs and adverse events in veterinary medicine.[14] This application of AI in comparative similarity association analysis enhances the ability to predict and mitigate potential adverse reactions in animal populations.

### 4.1.3 Clinical Trials

Clinical trials are a crucial phase in the drug development process, serving as a systematic approach to evaluate the safety and efficacy of new medical interventions. These trials are essential for determining whether a new drug or treatment is both effective and safe for end users. The importance of clinical trials lies in their role as a safeguard for public health, ensuring that only thoroughly tested and proven treatments reach the market. Clinical trials can be expensive, often costing millions to billions of dollars, depending on the drug's complexity, trial phase, number of participants, and duration. The cost also factors in expenses related to patient recruitment, administration, regulatory compliance, and data analysis. The integration of Artificial Intelligence (AI) into clinical trials is revolutionizing this process.

AI shows potential for being leveraged to improve patient recruitment strategies for certain areas. For instance, AI-driven tools are being developed to refine the quality of examination metrics, thus enhancing the precision of metric evaluations in clinical trials for conditions, for example, in inflammatory bowel disease.[15] These tools utilize AI to analyze imaging data, enabling better patient selection and monitoring throughout the trial process. The predictive capabilities of AI are also being harnessed to forecast clinical trial outcomes, as evidenced by research in the United Arab Emirates. Natural language processing (NLP) and multi-modal AI integrate disparate data sources, including patient demographics, clinical endpoints, and biomarker data, to predict the success rates of clinical trials.[16] This approach aids in the strategic planning of clinical trials by assessing the impact of target choice and trial design on outcomes.

AI is also optimizing eligibility screening processes for clinical trials. AI tools utilizing an NLP pipeline and protocol library has been developed to streamline the screening process. This ensures that the right candidates are selected for trials, increasing the likelihood of successful outcomes and reducing the risk of adverse events.[17] Recent research conducted by Italian scholars has elucidated the application of machine learning techniques, particularly the Random Forest algorithm and a comprehensive suite of other ML algorithms, in the context of phase I clinical trials. These algorithms help identify clinical-biomolecular markers that are significantly associated with toxicity, aiding in the early detection and mitigation of potential adverse reactions.[18]

Generative AI, especially when combined with artificial neural networks, is pioneering the concept of digital twins in drug development. Research in Germany suggests that these digital representations of patients, empowered by AI, can simulate individual responses to drugs, providing a personalized approach to trial simulations and potentially reducing the need for extensive human trials.[19] China's advancements in AI are contributing to the development of organ-on-a-chip technology. Reinforcement learning (RL) algorithms are integrated with these microfluidic chips to simulate human organ responses to various substances. This innovative approach could revolutionize drug evaluation by providing a more accurate and ethical alternative to traditional animal testing.[20]

## 4.2 Smart Manufacturing

The integration of Artificial Intelligence (AI) into the manufacturing processes of the pharmaceutical industry is a significant step towards enhancing both efficiency and quality. Taking a specific example, this section delves into the specifics of in silico prediction for process optimization and the role of predictive controllers in cell culture processes. AI extends beyond process optimization to include predictive maintenance, traceability of manufacturing processes.

In silico prediction, facilitated by machine learning pipelines, is a theoretical concept that stands to revolutionize drug screening and optimization. By simulating the drug development process within a virtual environment, machine learning algorithms can predict the interactions and stability of pharmaceutical compounds before they are physically manufactured.[23] This approach allows for the identification and correction of potential issues early in the development cycle, drastically reducing the time and resources typically required for drug screening.

USA scholars mentioned that machine learning-based models are being theorized to maximize the efficiency of cell culture processes, which are crucial for biopharmaceutical production. These predictive controllers are designed to regulate the conditions of cell cultures, such as nutrient supply, temperature, and pH levels, to ensure optimal cell growth and product yield. By continuously analyzing data from the culture environment, the machine learning model can make real-time adjustments to the process parameters.[24] This not only maximizes the growth and viability of the cell cultures but also ensures consistency and quality in the production of biological products. The application of such AI-driven predictive control systems in cell culture processes marks a significant advancement towards automated and precision-driven biomanufacturing.

The use of AI, particularly Density-Based Spatial Clustering of Applications with Noise (DBSCAN), is being applied to predictive maintenance of pharmaceutical equipment. This AI technique analyzes operational data to identify patterns that can predict equipment failure, allowing for timely maintenance that prevents costly downtime.[25] By anticipating maintenance needs, pharmaceutical plants can avoid unexpected breakdowns, thereby reducing the overall maintenance costs and increasing the longevity of their equipment.

Traceability in manufacturing is crucial for compliance with regulatory standards and ensuring product quality. A federated learning framework in Greece is being utilized to enhance traceability across the manufacturing process. This decentralized approach to machine learning enables multiple manufacturing units to collaborate and learn a shared prediction model while keeping their data localized, thus ensuring data privacy and security.[26] It's particularly effective in environments where data cannot be centralized due to privacy concerns or regulatory restrictions, allowing for traceability without compromising proprietary information.

As researched by Russia scholars, AI systems are theorized to bring about a significant transformation in energy management, particularly in optimizing the drying processes crucial for pharmaceutical production. An intelligent control and management information system can dynamically adjust the drying parameters in real-time, responding to data on humidity, temperature, and other critical factors that affect the drying process.[28] By optimizing these parameters, the AI system ensures efficient energy use, maintains product quality, and reduces waste, leading to cost-effective and sustainable manufacturing practices.

the Internet of Things (IoT) is being utilized in conjunction with AI to develop smart healthcare and safety management systems in India. IoT devices equipped with sensors collect vast amounts of health data, which AI systems then analyze to provide insights into patient health trends, predict potential health risks, and automate safety protocols. Such systems are especially beneficial in monitoring patient conditions in real-time, providing tailored healthcare solutions, and ensuring the safety of both patients and healthcare professionals.[29] The application of AI in IoT also extends to the management of healthcare facilities, optimizing operations and enhancing the quality of care provided.

## 4.3 Quality Control

Artificial Intelligence (AI) is significantly impacting the field of quality control (QC) and quality assurance (QA). This section reviews the deployment of machine vision and AI in ensuring the quality of pharmaceutical products, from coating thickness measurement to precision in chemotherapy compounding.

The integration of machine vision and deep learning for real-time measurement of coating thickness and defect detection in film-coated tablets is currently under research in Hungary. This method leverages machine learning (ML) algorithms to analyze visual data, accurately detect inconsistencies in the coating, and identify defects that could compromise the medication's efficacy or safety. The implementation of this technology in coating thickness measurement and defect recognition ensures that only tablets conforming to stringent quality standards are delivered to patients, upholding the high safety and efficacy standards of pharmaceutical products.[30]

Machine vision techniques, such as template matching and Haar cascades, have been utilized in Iran's industrial product lines for robust QC and quality assurance.[31] These ML-based approaches are effective for tasks like counting blister cards within drug packages, where precision is crucial. By automating the inspection process, these AI tools significantly reduce the chance of human error, increase the speed of the production line, and ensure that packaging is consistent and accurate. In France, a theoretical model for a video-assisted control system is proposed to enhance the safety and precision in chemotherapy compounding units.[32] AI algorithms would process video data to ensure that compounding procedures adhere to prescribed formulations, minimizing the risk of human error. This system could significantly impact patient safety by guaranteeing that chemotherapy treatments are prepared with exact specifications, thus improving treatment outcomes.

In Pharma pipelines, innovations also existed in defect detection as part of pharmaceutical quality assurance. In Singapore, machine learning technologies combined with web crawlers are being applied for product defect surveillance.[33] This application involves scanning various digital platforms to identify and report potential product defects mentioned online. Such a system allows pharmaceutical companies to quickly respond to any indications of product issues reported by consumers or healthcare professionals, thus maintaining high standards of product quality and patient safety.

Deep learning convolutional neural networks (CNNs) also been applied for the detection of internal defects in tablets.[34] CNNs are particularly adept at analyzing complex image data, allowing them to identify inconsistencies within tablets that may not be apparent to the human eye. This high-throughput, accurate, and adaptable approach to quality control ensures that defective tablets are identified and removed from the production line before they reach the consumer, thereby safeguarding the integrity of the pharmaceutical products. In China, a theoretical multi-sensor image fusion method is being explored to detect defects during the powder bed fusion (PBF) process, which is a type of 3D printing technology used for manufacturing drug delivery devices.[35] By combining data from multiple sensors, this method aims to create a comprehensive view of the PBF process, detecting any anomalies that could lead to defects in the final product. Early detection of such defects is crucial for preventing the wastage of materials and ensuring the reliability of the manufactured devices.

## 4.4 Regulatory Affairs

Regulatory affairs are a cornerstone of the pharmaceutical industry, serving as a vital link between pharmaceutical companies and regulatory bodies. This key sector focuses on ensuring that pharmaceutical products adhere to the extensive array of regulations and laws that govern the industry. This section explores the transformative impact of Artificial Intelligence (AI) in reshaping regulatory affairs. This includes its role in streamlining compliance data management, automating routine administrative tasks, and improving communications through advanced generative chat tools.

Central to this technological evolution is Document Intelligence, a key aspect of machine learning. The implementation of Language Learning Models (LLM) and Natural Language Processing (NLP) in the biomedical and regulatory domains has led to significant progress in areas such as document management, the extraction of pertinent information, and the harmonization of data. These technologies are proving to be indispensable for managing the extensive and complex documentation that is a staple of regulatory procedures. With AI and NLP, regulatory affairs are undergoing a major shift toward more efficient, precise, and streamlined processes, thus revolutionizing the way the pharmaceutical industry navigates the convoluted regulatory landscape. Additionally, NLP is making groundbreaking strides in the realm of healthcare communication. This is particularly evident in the development of chatbots capable of interacting with patients, offering a range of assistance in various medical scenarios. These advanced communication tools are not only enhancing patient engagement but also contributing to more efficient and accessible healthcare services.

India is witnessing the application of AI to facilitate digital transformation in pharmaceutical regulatory affairs.[36] This integration streamlines the regulatory process, making it more efficient and effective. AI systems can manage and analyze large volumes of regulatory data, track changes in regulatory guidelines, and help companies ensure that their products comply with the latest standards. Such applications of AI are pivotal in managing the complexities of global regulatory requirements, ultimately speeding up the time to market for new pharmaceutical products.

In Italy, AI is being utilized for automated methods of calculating, monitoring, and predicting compliance. This includes the application of analytics and blockchain technology to ensure data integrity within the pharmaceutical industry.[37] AI algorithms can detect anomalies in data that might indicate compliance issues, while blockchain provides an immutable record that enhances traceability and accountability in pharmaceutical manufacturing and distribution. Belgium scholars are exploring the use of multiple AI frameworks to automate routine work in pharmacovigilance.[38] AI-based static and dynamic systems are theorized to validate intelligent automation systems, drawing insights from good manufacturing practices. Such systems could automate the detection and reporting of adverse drug reactions, streamline the process of drug safety monitoring, and ensure that manufacturers promptly address potential safety issues.

The theoretical use of large language models in medicine is being explored, particularly through the deployment of LLM chatbots in biomedical contexts by UK researchers.[39] These chatbots are designed to understand and respond to complex medical queries, providing support to healthcare professionals and patients. By leveraging vast amounts of medical literature and patient data, LLM chatbots can offer insights, diagnose conditions, and even suggest treatments, thereby augmenting the capabilities of medical staff and improving patient care.

In the USA, a framework for document classification and information extraction has been developed for insurance applications using document intelligence and NLP.[40] This machine learning-based framework automates the categorization of documents and the extraction of key information, such as patient details and medical conditions from various types of documents. Such automation greatly enhances efficiency and accuracy in handling large volumes of paperwork, ensuring that important information is captured correctly and is easily accessible.

Sweden has applied NLP in a practical approach to harmonize product information (PI) across European medicinal products.[41] This application of NLP ensures consistency and clarity in the information provided to patients and healthcare providers, facilitating better understanding and safer use of pharmaceutical products. By processing diverse datasets and aligning content across multiple languages and formats, NLP aids regulatory bodies and pharmaceutical companies in maintaining standardized product information.

India researchers pointed out dynamic NLP-enabled chatbots are being deployed to improve healthcare access in rural areas. These chatbots are designed to interact with patients in a conversational manner, providing them with medical information, guidance on common health issues, and even preliminary diagnoses. By leveraging ML techniques, these chatbots can learn from each interaction, thereby continuously improving their ability to communicate effectively with patients.[42] This application of NLP is crucial in regions where healthcare resources are limited, offering an accessible and cost-effective solution for patient support.

A scoping review conducted in Qatar has focused on the technical metrics used to evaluate healthcare chatbots. This review assesses the effectiveness of chatbots by examining their accuracy, usability, and patient satisfaction. The application of NLP in healthcare chatbots aims to simulate human-like interactions, making medical advice more accessible and understandable to patients.[43] By establishing robust evaluation metrics, healthcare providers can ensure that chatbots meet the required standards for delivering reliable and helpful medical assistance. Turkey researcher has explored the potential of NLP-based chatbot platforms serving as consultants in oral surgery. These chatbots can provide surgeons with real-time information, suggest treatment plans based on current medical guidelines, and assist in patient education before and after surgery.[44] Such NLP applications in oral surgery demonstrate the potential for chatbots to support complex medical decision-making processes and enhance patient care.

## 4.5 Clinical and Hospital

Artificial Intelligence (AI) and Machine Learning (ML) are revolutionizing clinical decision support across the globe, aiding healthcare professionals in diagnosis, treatment planning, and patient management.

In the USA, the performance of a convolutional neural network (CNN) for interpreting 12-lead electrocardiograms (ECGs) has been compared to that of medical doctors. The CNN employs deep learning algorithms to analyze ECG data and has achieved comparable performance levels.[45] Moreover, explainability techniques are applied to make the AI's decision-making process transparent and understandable to healthcare providers. This advancement in AI offers significant support in cardiology, providing rapid and accurate ECG interpretation that can lead to timely and effective patient care.

Canada has developed the APPRAISE-AI tool for evaluating the methodological rigor and clinical utility of AI models. This tool utilizes frameworks such as MI-CLAIM, TRIPOD-AI, and STARD-AI to quantitatively assess AI studies that support clinical decision-making.[46] APPRAISE-AI ensures that AI applications in healthcare are both scientifically sound and practically useful, enhancing the reliability of AI tools used in clinical settings. China's research focuses on an ML-aided cross-reactive sensing array for cerebrospinal fluid (CSF) analysis. This intelligent clinical lab approach aids in diagnosing post-neurosurgical meningitis by analyzing CSF with greater accuracy and speed than traditional methods.[47] By harnessing ML algorithms, this technology can decipher complex patterns within CSF that are indicative of meningitis, facilitating early and accurate diagnoses.

In Malaysia, reinforcement learning models and algorithms are being applied to diabetes management. These models analyze patient data and make predictions and recommendations for treatment adjustments.[48] The use of reinforcement learning in this context allows for personalized diabetes care, as the algorithms adapt to the unique responses of each patient to treatment over time. GA-DRUG, a genetic algorithm-based ensemble learning framework for drug combination prediction has been stated.[49] This AI application identifies potentially effective drug combinations, which is critical in the treatment of complex diseases such as cancer. By predicting synergistic effects of drug combinations, GA-DRUG assists in optimizing treatment regimens and improving patient outcomes.

## 4.6 Computer Vision

Image recognition has taken a pivotal role in advancing diagnostic capabilities within the pharmaceutical industry, thanks to the evolution of AI, particularly in the form of deep learning neural networks.

In the realm of oncology, AI has made significant strides, particularly in the diagnosis of colorectal cancer through the analysis of histopathology images. Transfer-learned deep convolutional neural networks (CNNs) have been developed to distinguish between healthy and cancerous tissues with remarkable accuracy.[21] These AI systems are trained on vast datasets of labeled images, allowing them to learn the intricate patterns that signify the presence of cancer cells. The implementation of such AI tools assists pathologists by providing a second, highly accurate opinion, reducing diagnostic times, and potentially increasing the survival rates through earlier detection.

DeepCycle, a specialized deep neural network, represents another leap forward, demonstrating AI's capability to reconstruct cell cycle trajectories from unsegmented cell images. This application of convolutional neural networks is particularly significant in understanding the proliferation of cells, which is a crucial aspect in both cancer research and the development of various drugs.[22] By automating the analysis of cell cycle phases, DeepCycle assists biologists and pharmacologists in observing and interpreting cellular behaviors more efficiently, thus accelerating research and development in cell-based therapies.

AI could be used with visual function estimation. In Japan, deep learning techniques are being utilized to estimate visual function from ultra-widefield fundus images, specifically for patients with retinitis pigmentosa.[50] Deep learning algorithms can analyze these images to detect and quantify pathological features that correlate with visual function, offering a non-invasive method to evaluate the progression of the disease. This approach provides a rapid assessment tool for ophthalmologists, aiding in the management of retinal diseases and potentially improving the quality of life for patients by facilitating early intervention.

AI could be used with longitudinal Brain MRI prediction. In the USA, image-to-image translation through spatial-intensity transform (SIT) is a novel deep learning approach being theorized for predicting the progression of brain changes in longitudinal MRI studies.[51] This technique allows for the translation of medical images from one time point to another, predicting how a patient's brain anatomy might change over time. This has significant implications for the early detection and treatment of neurological conditions, as it can provide clinicians with a predictive look at disease progression, enabling more proactive management of conditions such as Alzheimer's disease or multiple sclerosis.

## 4.7 Other Supporting Area

In the pharmaceutical sector, technological advancements are significantly enhancing various supporting areas. Blockchain technology is being integrated into Supply Chain Management to optimize the medical supply chain, ensuring greater traceability and security. AI algorithms are instrumental in extracting medication information and identifying adverse drug events (ADEs) from clinical documents, thereby bolstering patient safety. Moreover, AI is revolutionizing laboratory practices by analyzing biological samples, such as urine, for faster and more accurate diagnostics. Large pharmaceutical laboratories also leverage AI and data analytics to navigate strategic decision-making, from R&D directions to market strategies, ensuring alignment with market needs and scientific progress. These innovations collectively represent the pharmaceutical industry's commitment to embracing technology for improved healthcare outcomes. The integration of Artificial Intelligence (AI) and blockchain technology is transforming the pharmaceutical industry, optimizing supply chain management and enhancing laboratory practices.

In India, the application of AI, specifically through deep learning, is being utilized to optimize supply chain management. The Archimedes Optimization with an Enhanced Deep Learning-based Recommendation System is designed to manage drug supply chains effectively.[52] This system can predict drug demand, streamline inventory management, and mitigate supply chain disruptions, ensuring efficient distribution of medications to where they are needed most. Blockchain technology is being explored for its potential to decentralize pharmaceutical supply chains. This theoretical framework promises to increase transparency, traceability, and security within the supply chain, as blockchain's immutable ledger would allow for the tracking of drugs from manufacture to delivery, reducing the risk of counterfeit medications entering the supply chain.[53] Named Entity Recognition (NER) and Relation Classification (RC) components are being applied to extract information about medications and associated adverse drug events (ADEs) from clinical documents.[54] This deep learning approach automates the extraction process, making it more efficient and reducing the likelihood of human error. It supports healthcare professionals by providing them with precise data for clinical decision-making and pharmacovigilance.

AI has show huge potential in laboratory management. Switzerland is advancing laboratory management with an AI system, utilizing artificial neural networks to analyze urine samples. The interpretable machine learning model, accessible via a GitHub repository, is capable of detecting chemically adulterated urine samples using high-resolution mass spectrometry.[55] This application not only enhances the accuracy of laboratory tests but also speeds up the process of sample analysis. The Mamdani model, a type of fuzzy logic system, is being theorized to assist with strategic decision-making in large pharmaceutical laboratories .[56] The model addresses the uncertainty and complexity of innovation risks, providing a framework for laboratories to make informed strategic choices that could affect their innovation pipeline and market competitiveness.

In India, various machine learning algorithms are employed for the prediction of pharmaceutical drug sales. These algorithms analyze historical sales data, market trends, and consumer behaviors to forecast future drug demands.[27] This information is crucial for manufacturers to adjust production levels, manage inventory, and plan for market fluctuations, ensuring that supply aligns with demand and minimizing the risk of overproduction or stock shortages.

# 5 Commercial Trend

This chapter comprehensively discusses AI startups active in the life and health sector, all of which have received investment from the venture capital firm Y Combinator. As one of the world's renowned VC firms, Y Combinator has recently paid particular attention to the biotechnology field. Their first investment in a biotech company, Ginkgo Bioworks, went public in 2021, demonstrating their keen investment insight and strength in this sector. The startups covered in this research are grouped into three categories: drug discovery an delivery, diagnostics and healthcare IT. By exploring the information, we not only gain improved comprehension about the innovative approaches created by these companies, but also have a more intuitive understanding about the investment preferences in the capital market 3.

There is a significant trend towards using advanced AI technologies to tackle complex problems in healthcare such as drug discovery, genomics therapies, mental health treatments, cancer diagnostics and medical data analysis. Developing treatments and diagnostics specially tailored to the unique needs of individual patients is increasingly becoming a central focus. Investment might flow towards startups that leverage AI to create more personalized healthcare solutions. With the growing digitization of healthcare data, there is a strong potential for investments in companies that offer innovative solution for data management, analysis and security, especially in the field of regulatory and compliance. Solutions that help healthcare providers and biotech companies navigate the complex regulatory landscape might become attractive investment opportunities. Moreover, the pandemic has accelerated the investment in remote health services, wearable health monitoring devices and platforms that facilitate virtual healthcare delivery.

Most of these companies were founded in the last few years which reflects the growing interest of investment in tech-driven healthcare solutions in recent years. The intersection of healthcare and technology is evidently a burgeoning field with substantial potential. It is likely that investment institutes will continue to heavily favor startups that integrate artificial intelligence into healthcare in the coming few years. However, it is important to point out that the preferences of investment institutes will also be influenced by broader economics trends, regulatory changes, technological advancements and global health challenges. Keeping an eye on those factors will be crucial for accurate predictions of future investment directions in the crossover sector of AI and healthcare.

# 6 Discussion

In the advent of the AI era, numerous traditional segments within the pharmaceutical industry are poised for transformative changes. A pertinent example is the field of regulatory affairs, which, despite its conventional roots, is now confronting an evolving landscape of challenges and opportunities in the AI epoch.

Table 3: Information about Startups invested by Y Combinator

| Category | Company Names | Year Founded | Main Business |
|---|---|---|---|
| drug discovery and delivery | Olio | 2023 | Olio labs uses AI to develop combination therapeutics that consider the thousands of interacting proteins in your body rather than targeting just one or two. |
| | WhiteLab Genomics | 2022 | WhiteLab Genomics focuses on accelerating the discovery and design of genomic therapies using AI. |
| | Mindstate Design Labs | 2021 | Mindstate Design Labs develops psychedelic therapeutics for mental health indications by combining machine learning, human experiential data, and molecular pharmacology. |
| | Helix Nanotechnologies | 2017 | Helix focuses on developing an AI-powered mRNA platform to address challenges related to cancer, COVID, and climate change. |
| Diagnostics | Cleancard | 2023 | Cleancard combines synthetic biology and AI to enable rapid identification of various cancers and biomarker tracking from urine samples. |
| | Scanbase | 2023 | Scanbase offers an AI-powered API that converts photos of rapid diagnostic tests into results. |
| | Segmed | 2020 | Segmed focuses on providing high-quality data for AI radiology by building the largest database of labeled medical data, starting with radiology images. |
| Healthcare IT | Cair Health | 2023 | Cair Health aims to automate medical insurance workflows for healthcare providers and payers using AI-powered software. |
| | Decoda Health | 2023 | Decoda Health utilizes AI-driven medical assistants to transform healthcare administration. |
| | Latent | 2023 | Latent builds medical language models to tackle operational overhead in healthcare. |
| | Invert | 2022 | Invert develops software to manage, analyze, and optimize bioprocessing data. By leveraging AI, Invert aims to enhance the efficiency and accuracy of bioprocessing. |
| | Delfino AI | 2022 | Delfino AI automates repetitive phone calls made by healthcare service providers provider offices to payors using artificial intelligence. |

Recent research highlights an increasingly complex environment in regulatory affairs, necessitating the adoption and integration of AI tools to navigate this dynamic domain.[57] This transition underscores the need for redefining traditional practices through innovative AI applications, ensuring efficient and effective management of regulatory processes.

AI's impact on the pharmaceutical industry is multifaceted, extending beyond enhancing efficiency and effectiveness. Its potential to reshape global healthcare is immense, as evidenced by its growing influence in both developed and developing nations. In the realm of pharmaceutical advancements, the judicious application of specific algorithms is crucial, particularly when considering their technological efficacy. The Random Forest algorithm, renowned for its ensemble learning capabilities, has shown exceptional effectiveness in drug discovery processes. Its proficiency in handling complex datasets with numerous variables makes it ideal for identifying drug-target interactions and assessing drug efficacies. Deep learning algorithms, particularly convolutional neural networks (CNNs), have set a benchmark in medical imaging analysis. Their advanced pattern recognition capabilities make them highly suitable for interpreting intricate medical imaging data, such as MRI and ECG scans. Their ability to learn from large volumes of data ensures high diagnostic accuracy, making them invaluable in the healthcare industry. In the domain of clinical data extraction, Natural Language Processing (NLP) algorithms, including traditional methods like Support Vector Machines (SVM) and advanced deep learning models like Transformers, have proven highly effective. These algorithms excel at extracting critical information from unstructured clinical texts, thereby enhancing data analysis efficiency and accuracy in pharmacovigilance. Moreover, the integration of blockchain algorithms in supply chain management marks a significant shift in the pharmaceutical industry. The technology's cryptographic algorithms and decentralized ledger systems ensure data integrity, traceability, and transparency, crucial for tracking drug distribution and preventing counterfeit medications. Lastly, neural networks are increasingly recognized for their potential in predictive analytics within pharmaceutical research. Their ability to process vast amounts of data and identify subtle patterns that might be overlooked by human researchers or simpler algorithms is invaluable. This capability is instrumental in predicting disease trends, patient outcomes, and market responses to new pharmaceutical products.

In the dynamic landscape of pharmaceutical development, the integration of Artificial Intelligence (AI) and Machine Learning (ML) stands as a beacon of transformation. This technological revolution, however, unfolds differently across the globe, with a pronounced disparity between developed and developing nations, each tailoring AI to meet their unique healthcare needs. In developing countries, exemplified by India, AI's swift adoption in key areas like supply chain management and healthcare delivery is not just a leap in technology but a strategic response to urgent regional challenges. For instance, AI-driven chatbots are being deployed in rural areas with limited healthcare infrastructure, providing basic medical guidance and preliminary diagnoses. AI is also redefining drug distribution networks in these regions, tackling logistical hurdles and infrastructure constraints by optimizing routes and forecasting supply needs, ensuring consistent access to medicines in even the most remote areas.

Contrastingly, developed nations are channeling AI to elevate precision and efficiency within their highly regulated healthcare systems. The United States epitomizes this trend through the adoption of deep learning for medical imaging analysis, including ECG and MRI scans. These AI-driven tools are enhancing diagnostic accuracy and laying the groundwork for predictive healthcare models, which preemptively identify potential health issues. This approach marks a significant stride towards personalized medicine, where treatments are customized to individual patient profiles. In Europe, with Switzerland and Spain as prime examples, there is a noticeable shift towards employing AI for complex tasks like automating laboratory operations and facilitating strategic decision-making in pharmaceutical companies. This reflects a broader strategy to leverage AI for bolstering research capabilities, ensuring quality control, and maintaining stringent compliance with pharmaceutical regulations.

This dichotomy in AI application between the developed and developing world illuminates their diverging strategic healthcare priorities. Developed countries are investing heavily in AI research and technology to redefine healthcare standards and protocols, whereas developing countries are applying AI solutions to bridge healthcare gaps, enhance service efficiency, and provide cost-effective solutions. This global trend underscores how developed countries are looking to AI for cutting-edge healthcare advancements, while developing countries utilize AI to make healthcare more accessible and reliable for their populations.

In the context of regulatory affairs, AI's role is crucial in navigating the complex maze of international standards, a task particularly relevant in developed countries with comprehensive pharmaceutical regulatory frameworks. Here, AI is instrumental in ensuring compliance, managing documentation, and streamlining new drug approval processes, thereby preserving the integrity of the pharmaceutical market and safeguarding public health. In developing countries, however, AI's focus in regulatory affairs leans towards maximizing accessibility and cost-effectiveness. Applications like blockchain technology in supply chain management ensure transparency and traceability in pharmaceutical products, while Natural Language Processing (NLP) systems facilitate the efficient extraction and processing of medical information, enhancing healthcare decision-making. In summary, the role of AI and ML in the pharmaceutical sector mirrors the varied priorities and challenges faced by different regions. The rapid application of AI in developing countries to tackle healthcare delivery challenges, contrasted with the focus on precision, innovation, and regulatory compliance in developed nations, highlights the versatile and transformative power of AI in addressing both global and region-specific pharmaceutical challenges.

As we look towards the future, the role of AI and ML in the pharmaceutical industry is anticipated to deepen and expand. In developing countries, AI is expected to become fundamental in democratizing healthcare access, significantly lowering barriers and providing scalable solutions for widespread medical issues. AI systems are likely to evolve beyond addressing basic healthcare needs, extending to manage more complex medical cases, particularly in regions currently underserved by traditional healthcare infrastructures. In developed countries, the future likely holds a refined focus on AI applications, aiming for greater precision in diagnostics and personalized medicine. The convergence of AI with genomics and biotechnology may lead to breakthroughs in targeted therapies and individualized treatment plans. As AI becomes more ingrained in these processes, ethical considerations, especially concerning data security and privacy, will gain prominence, necessitating evolved policies and regulations to ensure responsible and ethical utilization of AI's potential.

Globally, the synergy between AI, blockchain technology, and the Internet of Things (IoT) is poised to strengthen supply chains, enhancing their resilience against disruptions and improving distribution efficiency. Additionally, AI's role in drug discovery and development could revolutionize these processes, significantly reducing the time and cost associated with drug development and clinical trials.

Overall, the selection of an appropriate algorithm in pharmaceutical applications must be strategically aligned with the specific data types and task complexities at hand. This strategic alignment of algorithmic capabilities with the unique requirements of pharmaceutical applications is essential to fully realize the potential of technological advancements in this field.

## 7   Competing Interests

The authors declare no competing interests.

## 8   Declaration of generative AI and AI-assisted technologies in the writing process

During the preparation of this work the author(s) used Open AI ChatGPT4 in order to words/sentences polishing. After using this tool/service, the author(s) reviewed and edited the content as needed and take(s) full responsibility for the content of the publication.